\documentclass[twocolumn,prd,unsortedaddress,superscriptaddress,showpacs,a4paper 
,nofootinbib]{revtex4-1} 
\usepackage{epsfig} 
\usepackage{subfigure} 
 
\def\beq{\begin{equation}} 
\def\enq{\end{equation}} 
\def\beqa{\begin{eqnarray}} 
\def\enqa{\end{eqnarray}}

\def\Tr{\mbox{ Tr }} 
\def\qq{\lag\bar{q}q\rag} 
\def\uu{\lag\bar{u}u\rag} 
\def\dd{\lag\bar{d}d\rag}

\def\G3{\lag g^3G^3\rag} 
\def\pli{p^\prime} 
 
\def\la{\lambda}

\def\si{\sigma}

\def\al{\alpha}

\def\lb{\label} 
\def\nnb{\nonumber} 
\def\nn{\nonumber}

\newcommand{\rag}{\rangle} 
\newcommand{\lag}{\langle}

\def\MeV{\mbox{ MeV}} 
\def\GeV{\mbox{ GeV}} 

\begin{document} 
 
\title{\sc Radiative decay of the  X(3872) as a mixed molecule-charmonium  
state in QCD Sum Rules} 
\author{M.~Nielsen} 
\email{mnielsen@if.usp.br} 
\affiliation{Instituto de F\'{\i}sica, Universidade de S\~{a}o Paulo,  
C.P. 66318, 05389-970 S\~{a}o Paulo, SP, Brazil} 
\author{C.M.~Zanetti} 
\email{carina@if.usp.br} 
\affiliation{Instituto de F\'{\i}sica, Universidade de S\~{a}o Paulo,  
C.P. 66318, 05389-970 S\~{a}o Paulo, SP, Brazil} 
 
\begin{abstract} 
 
  We use QCD  sum rules to calculate the width  of the radiative decay 
  of the meson  $X(3872)$, assumed to be a  mixture between charmonium 
  and    exotic   molecular    $[c\bar{q}][q\bar{c}]$    states   with 
  $J^{PC}=1^{++}$.  We  find that in a  small range for  the values of 
  the  mixing  angle, $5^0\leq\theta\leq13^0$,  we  get the  branching 
  ratio             $\Gamma(X\to             J/\psi\gamma)/\Gamma(X\to 
  J/\psi\pi^+\pi^-)=0.19\pm0.13$,  which is  in agreement,  with the   
  experimental value.   This result  is compatible 
  with  the  analysis  of  the  mass  and  decay  width  of  the  mode 
  $J/\psi(n\pi)$ performed in the same approach. 
 
\end{abstract}

\pacs{ 11.55.Hx, 12.38.Lg , 12.39.-x} 
\maketitle 
 
%
%
\section{Introduction} 
%
%
The $X(3872)$ state has been first observed by the Belle collaboration 
in                              the                              decay 
$B^+\!\rightarrow\!X(3872)K^+\rightarrow\!J/\psi\pi^+\pi^-         K^+$ 
\cite{belle1},  and   was  later  confirmed  by  CDF,   D0  and  BaBar 
\cite{Xexpts}.      The    current     world    average     mass    is 
$m_X=(3871.4\pm0.6)\MeV$, and  the width is $\Gamma<2.3  \MeV$ at 90\% 
confidence level.  Babar  collaborations reported the radiative decay 
mode $X(3872)\to \gamma J/\psi$ \cite{belleE,babar2}, which determines 
$C=+$. Belle Collaboration reported the branching ratio: 
\beq            \frac{\Gamma(X\to\,J/\psi\gamma)}{\Gamma(X\to 
  J/\psi\,\pi^+\pi^-)}=0.14\pm0.05. \lb{data} \enq 
Further studies  from Belle and CDF  that combine  angular  
information and kinematic properties of the $\pi^+\pi^-$ pair,  
strongly favors the 
quantum numbers $J^{PC}=1^{++}$ or $2^{-+}$ \cite{belleE,cdf2,cdf3}.  
Between these quantum numbers, a recent BaBar measurement favors the  
$J^{PC}=2^{-+}$ assignment \cite{babarate}. However, established  
properties of the $X(3872)$ are in   
conflict with this assignment \cite{kane,bpps} and, therefore, in this 
work we assume the quantum numbers of the $X(3872)$ to be $J^{PC}=1^{++}$. 
 
The interest in this new state  has been increasing, since the mass of 
the $X(3872)$  could not be related  to any charmonium  state with the 
quantum  numbers  $J^{PC}=1^{++}$   in  the  constituent  quark  models 
\cite{bg}, indicating that  the conventional quark-antiquark structure 
should by  abandoned in  this case.  Another  interesting experimental 
finding is the fact that the decay rates of the processes $X(3872) \to 
J/\psi\,\pi^+\pi^-\pi^0$   and  $X(3872)\rightarrow\!J/\psi\pi^+\pi^-$ 
are      comparable     \cite{belleE}:     \beq      {\Gamma(X     \to 
  J/\psi\,\pi^+\pi^-\pi^0)\over                                 \Gamma( 
  X\to\!J/\psi\pi^+\pi^-)}=1.0\pm0.4\pm0.3.  \lb{rate}\enq 
This ratio indicates a strong isospin and G parity violation, which is 
incompatible with a $c\bar{c}$ structure for $X(3872)$. The decay  
$X \to J/\psi\omega$ was also observed by BaBar  
Collaboration \cite{babarate} at a rate: 
\beq 
{{\cal B}(X \to J/\psi\pi^+\pi^-\pi^0)\over {\cal B}(X\to\!J/\psi\pi^+ 
\pi^-)}=0.8\pm0.3, 
\label{barate} 
\enq 
which is consistent with the result in Eq.~(\ref{rate}).

The isospin  violation problem  can be easily  avoided in  a multiquark 
approach.  In this context the molecular picture has gained attention. 
The observation  of the above  mentioned decays, plus  the coincidence 
between    the    $X$   mass    and    the   $D^{*0}D^0$    threshold: 
$M(D^{*0}D^0)=(3871.81\pm0.36)\MeV$ \cite{cleo}, inspired the proposal 
that      the      $X(3872)$       could      be      a      molecular 
$(D^{*0}\bar{D}^0-\bar{D}^{*0}D^0)$  bound  state  with small  binding 
energy \cite{close,swanson}.  The $D^{*0}\bar{D}^0$ molecule is not an 
isospin eigenstate and the rate in Eq.~(\ref{rate}) could be explained 
in a very natural way in this model.  
 
Although  the  molecular picture  is  gaining  attention with  studies 
indicating that  it can  be a suitable  description for  the $X(3872)$ 
structure \cite{nnl}, there are  also some experimental data that seem 
to indicate the existence of  a $c\bar{c}$ component in its structure. 
In ref.~\cite{Bignamini:2009sk}, a simulation for the production of a bound  
$D^0\bar{D}^{*0}$ state with biding energy as small as 0.25 MeV, reported a 
production cross section that is an order of magnitude smaller than the 
cross section obtained from the CDF data. A similar 
result was obtained in ref.~\cite{suzuki} in a more phenomenological 
analysis. However, as pointed out in ref.~\cite{Artoisenet:2009wk}, a  
consistent analysis of the $D^0\bar{D}^{*0}$ molecule production requires  
taking into account the effect of final state interactions of the $D$ and  
$D^*$ mesons.  
 
Besides this debate, the  recent observation,  reported by 
BaBar \cite{babar09},  of the  decay $X(3872)\to \psi(2S)\gamma$  at a 
rate:    \beq   {{\cal    B}(X   \to    \psi(2S)\,\gamma)\over   {\cal 
    B}(X\to\psi\gamma)}=3.4\pm1.4, 
\label{rategaexp} 
\enq 
is much bigger than the molecular prediction ~\cite{swan1}: 
\beq 
{\Gamma(X \to \psi(2S)\,\gamma)\over\Gamma(X\to\psi\gamma)}\sim4\times 
10^{-3}. 
\label{ratega} 
\enq 
 
Another interesting interpretation for the $X(3872)$ is that it 
could be a compact tetraquark state \cite{maiani,tera,tera2,x3872}. In  
particular, Terasaki ~\cite{tera2} argues that with a tetraquark  
interpretation the ratio in Eq.~(\ref{data}) could be easily explained.   
 
In Ref.\cite{x3872mix}  the QCDSR  approach was used  to study  the $X$ 
structure including the  possibility  of the mixing between two and 
four-quark  states.  This was  implemented following  the prescription 
suggested in \cite{oka24} for the light sector.  The mixing is done at 
the level of  the currents and is extended to the  charm sector.  In a 
different  context (not  in  QCDSR), a  similar  mixing was  suggested 
already  some  time ago  by  Suzuki  \cite{suzuki}.  Physically,  this 
corresponds to  a fluctuation  of the $c  \overline{c}$ state  where a 
gluon is emitted and  subsequently splits into a light quark-antiquark 
pair,  which lives  for some  time  and behaves  like a  molecule-like 
state.  The possibility that the $X(3872)$ is the mixing of two-quarks 
and molecular states was also considered to investigate the radiative decay 
in the effective Lagrangian approach \cite{dong08}, and to explain the  
data from BaBar \cite{babar09} and Belle \cite{belle08} using a  
Flatt\'e analysis \cite{kane2}. 
 
In  this  work  we  will  focus on  the  radiative  decay  $X(3872)\to 
J/\psi\gamma$. We use the  mixed two-quark and four-quark prescription 
of Ref.\cite{x3872mix} 
to perform a QCD sum  rule analysis of the radiative decay $X(3872)\to 
J/\psi\,\gamma$.  
%
%
\section{The mixed two-quark / four quark operator} 
%
%

The mixed  charmonium-molecular current proposed  in Ref.\cite{x3872mix} 
will be used to study radiative  decay of the $X(3872)$ in the QCD sum 
rules framework. 
 
For the charmonium part we use the conventional axial current: 
\beq  j'^{(2)}_{\mu}(x) = \bar{c}_a(x)  \gamma_{\mu} \gamma_5 
c_a(x).  \lb{curr2} \enq 
 
The $D$ $D^{*}$ molecule is interpolated by \cite{liuliu,dong,stancu}: 
\beqa 
j^{(4q)}_{\mu}(x) & = & {1 \over \sqrt{2}} 
\bigg[ 
\left(\bar{q}_a(x) \gamma_{5} c_a(x) 
\bar{c}_b(x) \gamma_{\mu}  q_b(x)\right) \nonumber \\ 
& - & 
\left(\bar{q}_a(x) \gamma_{\mu} c_a(x) 
\bar{c}_b(x) \gamma_{5}  q_b(x)\right) 
\bigg], 
\lb{curr4} 
\enqa 
As in Ref.~\cite{oka24} we define the normalized two-quark current as 
\beq 
j^{(2q)}_{\mu} = {1 \over 6 \sqrt{2}} \uu j'^{(2)}_{\mu},\lb{j2q} 
\enq 
and from these two currents we build the following  mixed  
charmonium-molecular current for the $X(3872)$: 
\beq 
J_{\mu}^q(x)= \sin(\theta) j^{(4q)}_{\mu}(x) + \cos(\theta)  
j^{(2q)}_{\mu}(x).  
\lb{field} 
\enq 
 
Following Ref.~\cite{x3872mix}  we will  consider a $D^0\bar{D}^{*0}$ 
molecular state with a small  admixture of 
$D^+D^{*-}$  and $D^-D^{*+}$  components: 
\beq    j_{\mu}^X(x)=    \cos\alpha 
J_{\mu}^u(x)+\sin\alpha J_{\mu}^d(x), 
\label{4mix} 
\enq 
with   $J_{\mu}^q(x)$, ($q=u,d$),  given  by   the   mixed 
two-quark/four-quark current in Eq.~(\ref{field}).

%
%
\section{The three point correlator} 
%
%
 
In this section we use QCD sum rules to study the vertex associated to 
the  decay  $X(3872)\to J/\psi\gamma$.   The  QCD  sum rules  approach 
\cite{svz,rry,SNB} is based on  the principle of duality.  It consists 
in the assumption that a correlation function may be described at both 
quark and hadron levels.   At the hadronic level (the phenomenological 
side)  the  correlation  function  is  calculated  introducing  hadron 
characteristics such  as masses and coupling constants.   At the quark 
level, the  correlation function is written  in terms  of quark and 
gluon fields and  a Wilson's operator product expansion  (OPE) is used 
to deal with the complex structure of the QCD vacuum. 
 
The   QCD   sum   rule   calculation  for   the   vertex   $X(3872)\,J/ 
\psi\,\gamma$ is centered around the three-point function given by 
\beq 
\Pi_{\mu\nu\al}(p,\pli,q)=\int d^4x d^4y ~e^{i\pli.x}~e^{iq.y} 
\Pi_{\mu\nu\al}(x,y),\lb{3p} 
\enq 
with 
\beq 
\Pi_{\mu\nu\al}(x,y)=\lag 0 |T[j_\mu^{\psi}(x)j_{\nu}^{\gamma}(y) 
{j_\al^X}^\dagger(0)]|0\rag, 
\lb{3po} 
\enq 
where $p=\pli+q$ and the interpolating fields  
are given by: 
\beq 
j_{\mu}^{\psi}=\bar{c}_a\gamma_\mu c_a, 
\lb{psi} 
\enq 
\beq 
j_{\nu}^{\gamma}=\sum_{q=u,d,c} e_q\,\bar{q}\gamma_\nu q\,, 
\lb{gamma} 
\enq 
with $e_q=\frac{2}{3}e$ for quarks $u$ and $c$, and $e_q=-\frac{1}{3}e$  
for quark $d$  ($e$ is the modulus  of the electron  charge).  The  
current $J_\mu^X$   is  given   by  the   mixed   charmonium-molecule   
current in Eq.~(\ref{4mix}).

In our analysis, we consider the  quarks $u$ and $d$ to be degenerate, 
{\it  i.e.},  $m_u=m_d$  and  $\uu=\dd=\qq$,  then  by  inserting  the mixed  
current (\ref{4mix}) in  Eq.~(\ref{3po}), we arrive at the following 
relation for the correlator 
\beqa 
\Pi_{\mu\nu\al}(x,y)&=&\frac{e\sin\theta}{3}(2\cos\alpha-\sin\alpha) 
\Pi_{\mu\nu\al}^{mol}(x,y)\nn\\ & + &{e\qq\over6\sqrt{2}}\cos\theta( 
\cos\al+\sin\al)\Pi^{c\bar{c}}_{\mu\nu\al}(x,y)\,.\nn\\ 
\label{sepi} 
\enqa 
The relation  for the  correlator is written  in terms of 
the charmonium and molecule  contributions. For the charmonium term we 
have 
\beq 
\Pi^{c\bar{c}}_{\mu\nu\al}(x,y)=\lag 0 |T[j_\mu^{\psi}(x)j_{\nu}^{\gamma}(y) 
{j_\al^{'(2)}}^\dagger(0)]|0\rag,\lb{corrcc} 
\enq 
and the molecular term is given by 
\beq 
\Pi^{mol}_{\mu\nu\al}(x,y)=\lag 0 |T[j_\mu^{\psi}(x)j_{\nu}^{\gamma}(y) 
{j_\al^{(4q)}}^\dagger(0)]|0\rag,\lb{corrmol} 
\enq 
with ${j_\al^{'(2)}}$ and ${j_\al^{(4q)}}$ given by Eqs.~(\ref{curr2}) and  
(\ref{curr4}) respectively.  
 
We now  proceed to  the calculation of  both charmonium  and molecular 
contributions in  in the OPE side.   By inserting the  currents of the  
two-quark component, $J/\psi$, and photon, respectively defined 
in    Eqs.~(\ref{j2q}),    (\ref{psi})   and    (\ref{gamma}),   in 
Eq.(\ref{corrcc}),  we  obtain  for  the charmonium  contribution  the 
following relation: 
\beqa 
\Pi^{c\bar{c}}_{\mu\nu\al}(x,y)&=&-{2\over3}\Tr\bigl[\gamma_\mu S^c_{ab}(x-y) 
\gamma_\nu S_{bc}^c(y)\gamma_\al\gamma_5S^c_{ca}(-x)\nn\\ 
&+&\gamma_\mu S^c_{ac}(x)\gamma_\al\gamma_5 S_{cb}^c(-y)\gamma_\nu  
S^c_{ba}(-x+y)\bigr]\,, 
\label{decc} 
\enqa 
 
\begin{center}  
\begin{figure*}[t] 
\includegraphics[height=30mm]{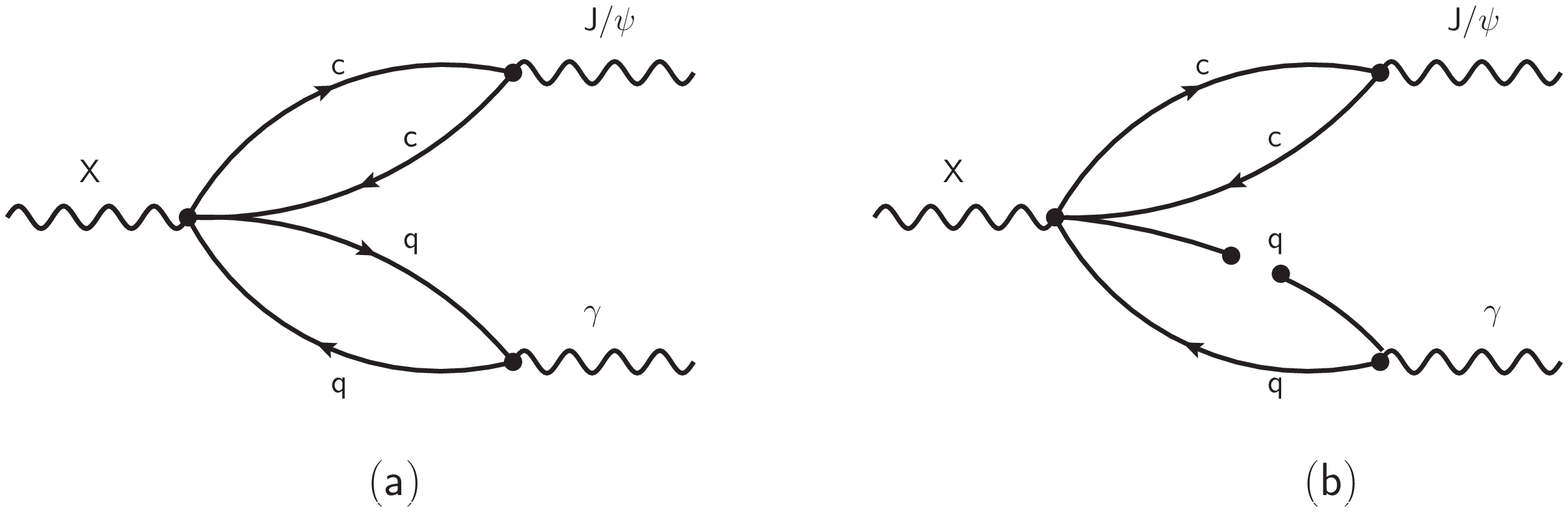} 
\includegraphics[height=30mm]{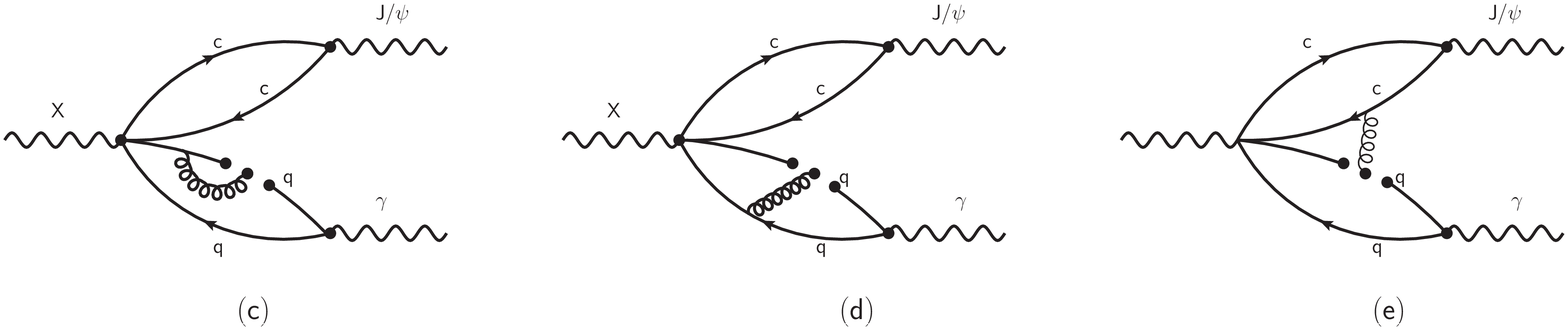} 
\includegraphics[height=25mm]{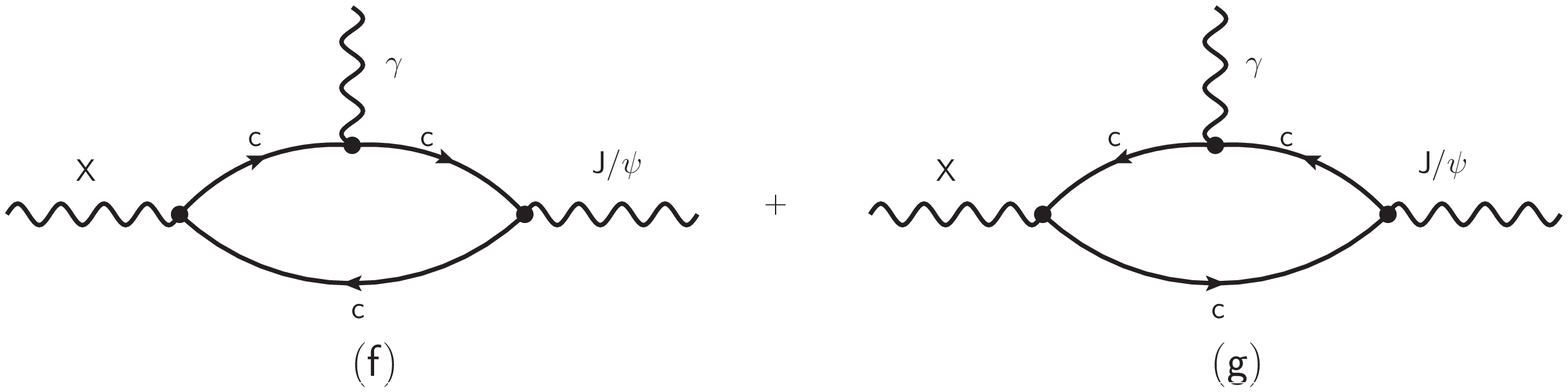} 
\caption{\label{3pdiags} Diagrams which contribute  to the OPE side of 
  the sum rule. Diagrams (a) to (e) contribute to the molecule term of 
  the OPE; (f) and (g) contribute to the charmonium term.} 
\end{figure*}  
\end{center}   
 
 
%
\noindent where $S^q_{ab}(x-y)=\langle0\vert T[q_a(x)\bar{q}_b(y)]\vert0\rangle$ 
is the full propagator of the quark $q$ (here $a,b,c$ are color indices). 
 
For the molecular contribution we use the four-quark current defined in 
Eq.~(\ref{curr4}), as  well as the  currents for the $J/\psi$  and the 
photon. Inserting these currents in Eq.~(\ref{corrmol}), we get 
\beqa 
&&\Pi_{\mu\nu\al}^{OPE}(x,y)=\frac{1}{\sqrt{2}}\Tr\biggl[\gamma_\mu  
S^c_{a'a}(x)\gamma_5S^q_{ab'}(-y)\gamma_\nu\times\nn\\&&\times S^q_{b'b} 
(y)\gamma_\al S^c_{ba'}(-x)-\gamma_\mu S^c_{a'c}(k)\gamma_\al S^q_{ab'}(-y) 
\gamma_\nu\times\nn\\&&\times S^q_{b'b}(y) 
\gamma_5 S^c_{ba'}(k-p')\biggr].\nn\\ 
\enqa 
 
To evaluate  the phenomenological side of  the sum rule  we insert, in 
Eq.(\ref{3p}), intermediate  states for $X$ and $J/\psi$.  We use the 
following definitions: 
\beqa 
&&\langle0\vert j_\mu^\psi\vert\psi(\pli)\rangle=m_\psi f_\psi 
\epsilon_\mu(\pli)\,;\\ 
&&\langle X(p)\vert j_\al^X\vert0\rangle=(\cos\alpha+\sin\alpha) 
\la_q\epsilon_\al^*(p)\,, 
\enqa 
where  the  meson-current coupling  parameter  is  extracted from  the 
two-point    function,    and    its    value    was    obtained    in 
Ref.~\cite{x3872mix}:  $\lambda_q = (3.6 \pm0.9) \times 10^{-3}  \GeV^5 $.   
We obtain the following expression: 
\beqa 
\Pi_{\mu\nu\al}^{\mathrm{phen}} (p,\pli,q)&=&-\frac{(\cos\alpha+\sin\alpha) 
\lambda_q m_{\psi}f_{\psi}\epsilon_\mu(\pli)\epsilon_\al^*(p)} 
{(p^2-m_{X}^2)({\pli}^2-m_{\psi})} 
\nn\\ 
&\times&\langle\psi(\pli)\vert j_\nu^\gamma\vert X(p)\rangle 
\,. 
\lb{phen} 
\enqa 
The  remaining  matrix   element   can be  related  to  the one  that   
describes the  decay 
$X\to\gamma J/\psi$: 
 
\beqa\langle\psi(\pli)\vert 
j_\nu^\gamma(q)\vert X(p)\rangle=i\,\epsilon_\nu^\gamma(q)\, 
\mathcal{M}(X(p)\to\gamma(q)J/\psi(\pli))\,,\nn\\\enqa 

\noindent and we can define \cite{dong08} 
\beqa 
&&\mathcal{M}(X(p)\to\gamma(q)J/\psi(\pli))= e\,\varepsilon^{\kappa 
\lambda\rho\sigma}\epsilon_X^\alpha(p)\epsilon^\mu_\psi(p^\prime) 
\epsilon^\rho_\gamma(q)\times\nn\\& &\times \frac{q_\sigma}{m_X^2}(A\, 
g_{\mu\lambda}g_{\al\kappa}p\cdot q+B g_{\mu\lambda}p_\kappa q_\al+C  
g_{\al\kappa}p_\lambda q_\mu),\lb{matrix} 
\enqa 
where  $A,B,C$ are  dimensionless couplings.   Using this  relation in 
Eq.(\ref{phen}),  we can write  the phenomenological  side of  the sum 
rule as: 
 \beqa  &&\Pi_{\mu\nu\al}^{\mathrm{phen}}   (p,\pli,q)=\frac{i  e 
(\cos\alpha+\sin\alpha)  \lambda_q m_{\psi}f_{\psi}}{m_X^2(p^2-m_{X}^2) 
({\pli}^2-m_{\psi})} 
\nn\\ 
&\times&\bigg(\epsilon^{\al\mu\nu\si}q_\si\,p\cdot q 
A+\epsilon^{\mu\nu\la\si}\pli_\la        q_\si       q_\alpha       B- 
\epsilon^{\al\nu\la\si}q_\mu q_\si\pli_\la C 
\nn\\ 
&+&\epsilon^{\al\nu\la\si}\pli_\la\pli_\mu      q_\si(C-A)\frac{p\cdot 
  q}{m_\psi^2} 
\nn\\ 
&                                     -&\epsilon^{\mu\nu\la\si}\pli_\la 
q_\si(q_\al+\pli_\al)(A+B)\frac{p\cdot  q}{m_X^2}\bigg)\,.  \lb{phenmix} 
\enqa 

In the  OPE side we work in  leading order in $\al_s$  and we consider 
condensates up to dimension  five, as shown in Fig.~\ref{3pdiags}.  In 
the phenomenological side, as we can see in Eq.~(\ref{phenmix}), there 
are five  independent structures.  We choose  one convenient structure 
to determine each one of  the couplings $A,B,C$ in Eq. (\ref{matrix}). 
Taking  the  limit  $p^2={\pli}^2=-P^2$   and  doing  a  single  Borel 
transform to $P^2\rightarrow M^2$, we  arrive at a general formula for 
the sum rule for each structure $i$: 
\beqa 
&&G_i(Q^2)\left(e^{-m_\psi^2/M^2}-e^{-m_X^2/M^2}\right)+H_i(Q^2)~ 
e^{-s_0/M^2}=\nn\\ &&=\bar{\Pi}_i^{(OPE)}(M^2,Q^2), 
\label{3sr} 
\enqa 
where  $Q^2=-q^2$  and  $H_i(Q^2)$   gives  the  contribution  of  the 
pole-continuum transitions \cite{decayx,dsdpi,io2}.  In the following, 
we show the expression of the  sum rules for the three structures that 
we have chosen to work.

\subsubsection{Structure 1: $\epsilon^{\al\mu\nu\si}q_\si$} 
 
The     RHS    of     the     sum    rule     for    the     structure 
$\epsilon^{\al\mu\nu\si}q_\si$ (structure 1) have both charmonium and 
molecule contributions: 
\beqa 
&&\bar{\Pi}_1^{\mathrm{OPE}}(M^2,Q^2)=-{\qq}\bigg[\frac{\sin\theta 
(2\cos\al-\sin\al)}{3Q^4}\times\nn\\&&\times\bar{\Pi}^{4q}_1(M^2,Q^2) 
+\frac{\cos\theta}{2Q^2}(\cos\al+\sin\al)\bar{\Pi}^{\bar{c}c}_1(M^2,Q^2) 
\bigg],\nn\\ 
\label{3sra} 
\enqa 
where the molecular contribution is given by 
\beqa 
\bar{\Pi}^{mol}_1(M^2,Q^2)&=&\bigg(1-{m_0^2\over3Q^2}\bigg) 
\int_{4m_c^2}^{u_0}du~e^{-u/M^2}u\times\nn\\ 
&\times&\sqrt{1-\frac{4m_c^2}{u}}\left({1\over2}+{m_c^2\over u}\right) 
+\nn\\ 
&+&{m_c^2m_0^2\over16}\int_0^1 d\al\frac{1+3\al}{\al^2(1-\al)}~e^{- 
\frac{m_c^2}{\al(1-\al)M^2}}\,.\nn\\ 
\label{est1} 
\enqa 
and the charmonium contribution is 
\beqa 
\bar{\Pi}_1^{\bar{c}{c}}(M^2,Q^2)&=& -\int^{s_0}_{4m_c^2}ds 
\int^{u_{+}}_{u_{-}}du\,{e^{-\frac{u+s}{M^2}}}\frac{2}{\sqrt{\la}} 
\times\nn\\&\times&\bigg(m_c^2+\frac{tu(t-u)}{\la}\bigg)\,,\nn\\ 
\lb{2q1} 
\enqa 
where $\la=\la(s,t,u)=s^2+t^2+u^2-2st-2su-2tu$, and $t=-Q^2<0$. 
 
In  the  above expressions  the  parameters $s_0=(m_X+\Delta_s)^2$  and 
$u_0=(m_\psi+\Delta_u)^2$  are the  continuum thresholds  for  $X$ and 
$J/\psi$ respectively. The limits of the integral in $u$ are: 
     \beq 
u_{\pm}=s+t+\frac{1}{2m_c^2}\left(-st\pm\sqrt{st(s-4m_c^2)(t-4m_c^2)} 
\right)\,. \enq 
The integrals in $s$ and $u$ also obey the following conditions:  
\beq 
 t<u,\,\;\;4m_c^2\leq s_0\,.  
\enq 

Since the  photon is  off-shell in the  vertex $X J/\psi\gamma$  it is 
required the introduction of form  factors. Then in the left hand side 
of the sum  rule, we define the function  $G_1(Q^2)$, which is related 
to the form factor $A(Q^2)$ as: 
       \beq        
G_1(Q^2)=\frac{3\sqrt{2}\pi^2(\cos\al+\sin\al)\la_qm_\psi 
  f_\psi}{m_X^2(m_X^2-m_\psi^2)}A(Q^2)\,. 
\label{Aq2} 
\enq  

\subsubsection{Structure 2: $\epsilon^{\mu\nu\si\la}\pli_\si\pli_\al q_\la$}

The     RHS    of     the     sum    rule     for    the     structure 
$\epsilon^{\mu\nu\si\la}\pli_\si\pli_\al q_\la$ (structure 2) has only 
molecular contribution: 
\beqa 
\bar{\Pi}^\mathrm{OPE}_2(M^2,Q^2)=\frac{m_0^2\qq}{Q^4}\int_0^1d\al 
\frac{1-\al}{\al}~e^{-\frac{m_c^2}{\al(1-\al)M^2}}\,. 
\label{3srb} 
\enqa 
 
In  the  left  hand side  of  the  sum  rule  we define  the  function 
$G_2(Q^2)$, which is related to the sum of form factor $A(Q^2)+B(Q^2)$ 
as: 
\beq 
G_2(Q^2)=\frac{3^22^4\sqrt{2}\pi^2(\cos\al+\sin\al)\la_qm_\psi  
f_\psi(A(Q^2)+B(Q^2))}{\sin\theta(2\cos\al-\sin\al)m_X^4(m_X^2-m_\psi^2)}\,. 
\label{Bq2} 
\enq 

\subsubsection{Structure 3: $\epsilon^{\al\nu\la\si}\pli_\la q_\si q_\mu$} 
 
The     RHS    of     the     sum    rule     for    the     structure 
$\epsilon^{\al\nu\la\si}\pli_\la q_\si q_\mu$  (structure 3) has only 
charmonium contribution: 
\beqa 
&&\bar{\Pi}_3^\mathrm{OPE}(M^2,Q^2)={\qq}\int^{s_0}_{4m_c^2}ds 
\int^{u_{+}}_{u_{-}}du\,{e^{-\frac{u+s}{M^2}}}\frac{2}{\la^{3/2}}\times 
\nn\\ &\times& \bigg[tu+m_c^2(-s+t+u)+\frac{3~s~t~u(-s+t+u)}{\la}\bigg]\,. 
\label{3src} 
\enqa  
The integrals  in this  equation  obey the  same relations  and 
conditions defined for the Eq.~(\ref{2q1}). 
 
In  the  left  hand side  of  the  sum  rule  we define  the  function 
$G_3(Q^2)$, which is related to the form factor $C(Q^2)$ 
as: 
\beq 
G_3(Q^2)=\frac{6\sqrt{2}\pi^2\la_qm_\psi f_\psi}{\cos\theta m_X^2 
(m_X^2-m_\psi^2)}C(Q^2)\,. 
\label{Cq2} 
\enq

\begin{figure*}  
 \subfigure[]{\label{fig2a}\includegraphics[width=0.5\textwidth]{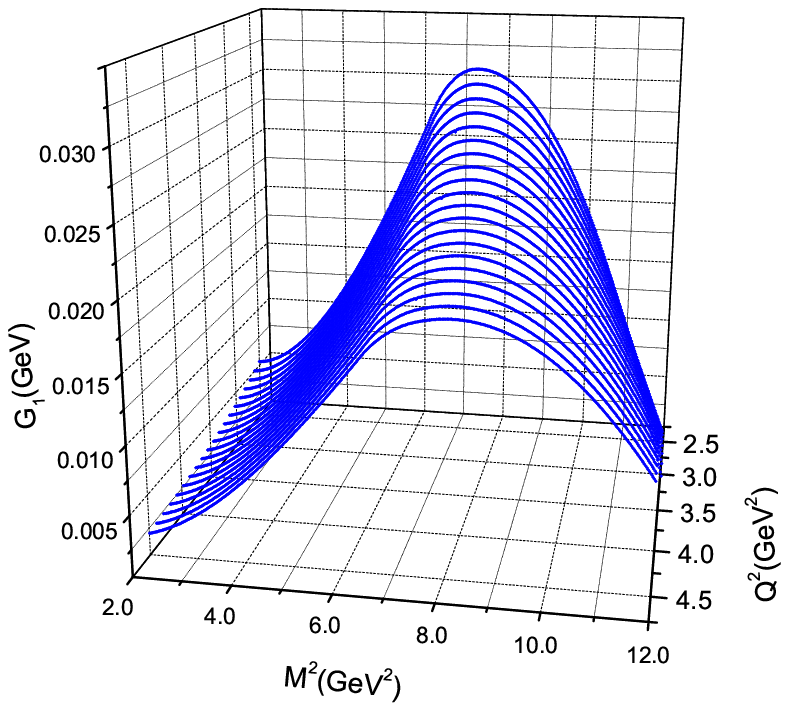}}\subfigure[]{\label{fig2b}\includegraphics[width=0.5\textwidth]{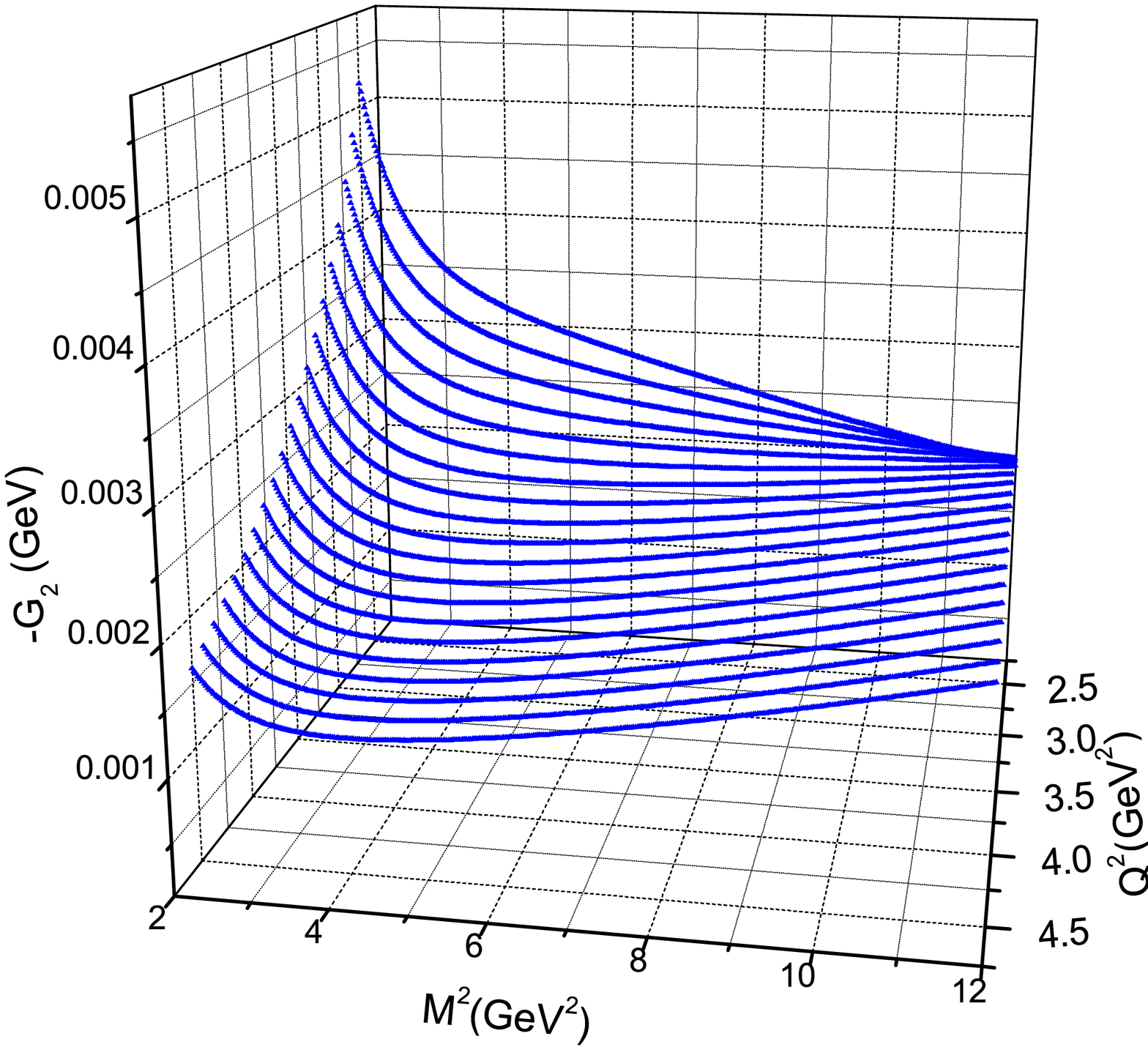}} 
  \subfigure[]{\label{fig2c}\includegraphics[width=0.5\textwidth]{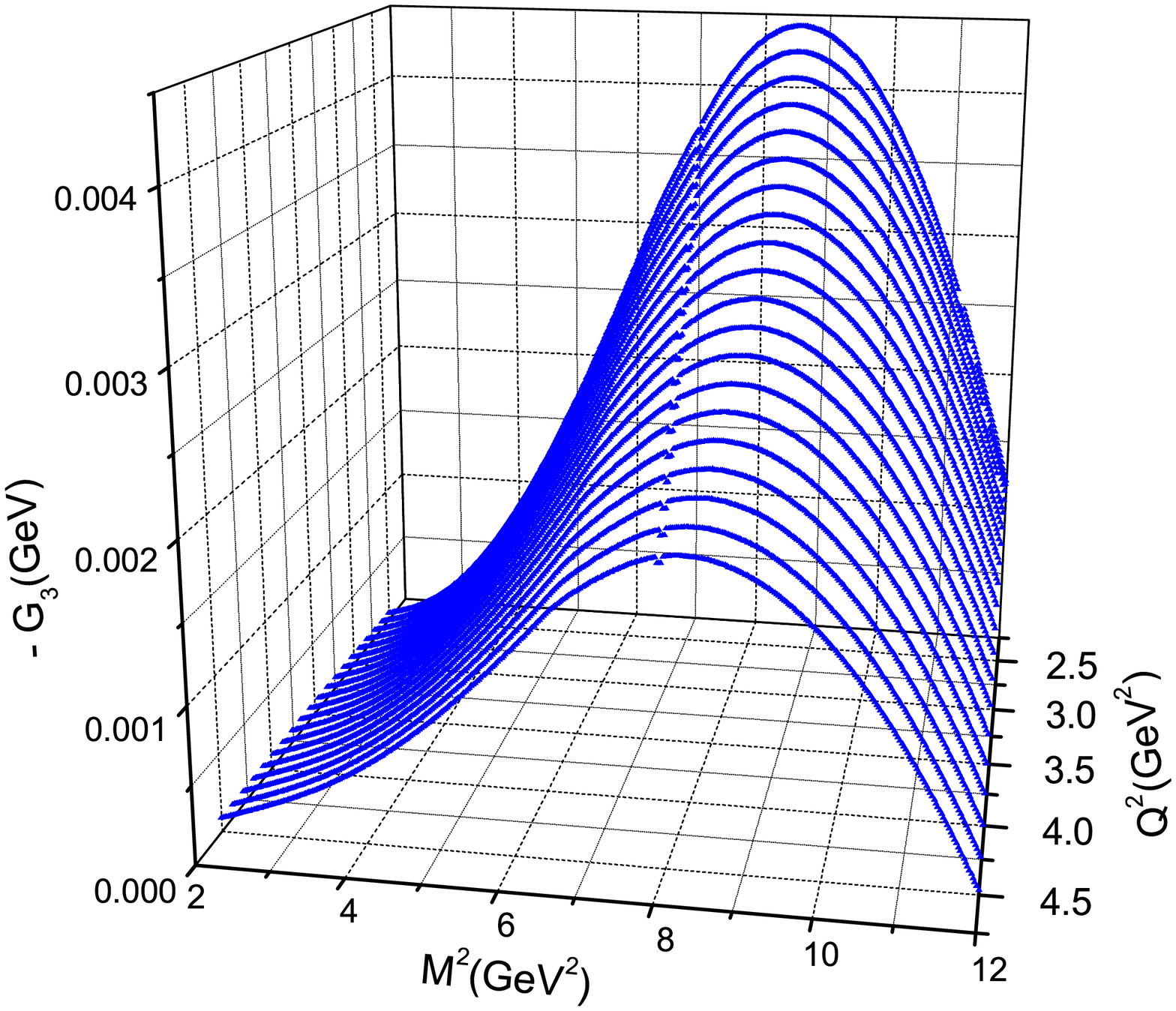}} 
\caption{\lb{fig2} Values of the functions obtained by varying both $Q^2$  
and $M^2$: a) $G_1(Q^2)$, b) $G_2(Q^2)$ and c) $G_3(Q^2)$.} 
\end{figure*} 
 
\section{Numerical analysis} 
 
The sum rules are analysed  numerically using the following values for 
quark masses  and QCD  condensates \cite{x3872,narpdg}, and  for meson 
masses e decay constants: 
\beqa\label{qcdparam} 
&m_c(m_c)=(1.23\pm 0.05)\,\GeV,\nnb\\ 
&\qq=-(0.23\pm0.03)^3\,\GeV^3,\nnb\\ 
&m_0^2=0.8\,\GeV^2,\nnb\\ 
&m_{\psi} = 3.1 \GeV\nnb\\, 
&m_{X} =3.87 \GeV\nnb\\ 
&f_{\psi} = 0.405 \GeV  
\enqa 

The  value of  the angle  $\al$ that  defines the  mixing  between the 
$D^0\bar{D}^{*0},~\bar{D}^0{D}^{*0}$  and  $D^+D^{*-},~D^-D^{*+}$  has 
been obtained previously in Ref.~\cite{decayx,x3872mix,maiani}: 
\beq 
\al=20^o 
\enq 
For the  mixing angle of two  and four quark states,  $\theta$, we use 
the values  that were obtained  in the QCD  sum rules analysis  of the 
mass   of   the  $X$   and   the   decay   mode  $X\to   J/\psi(n\pi)$ 
\cite{x3872mix}: 
\beq \theta=(9\pm4)^o.\enq 
%
 
\begin{figure*} 
  \subfigure[]{\includegraphics[width=0.45\textwidth]{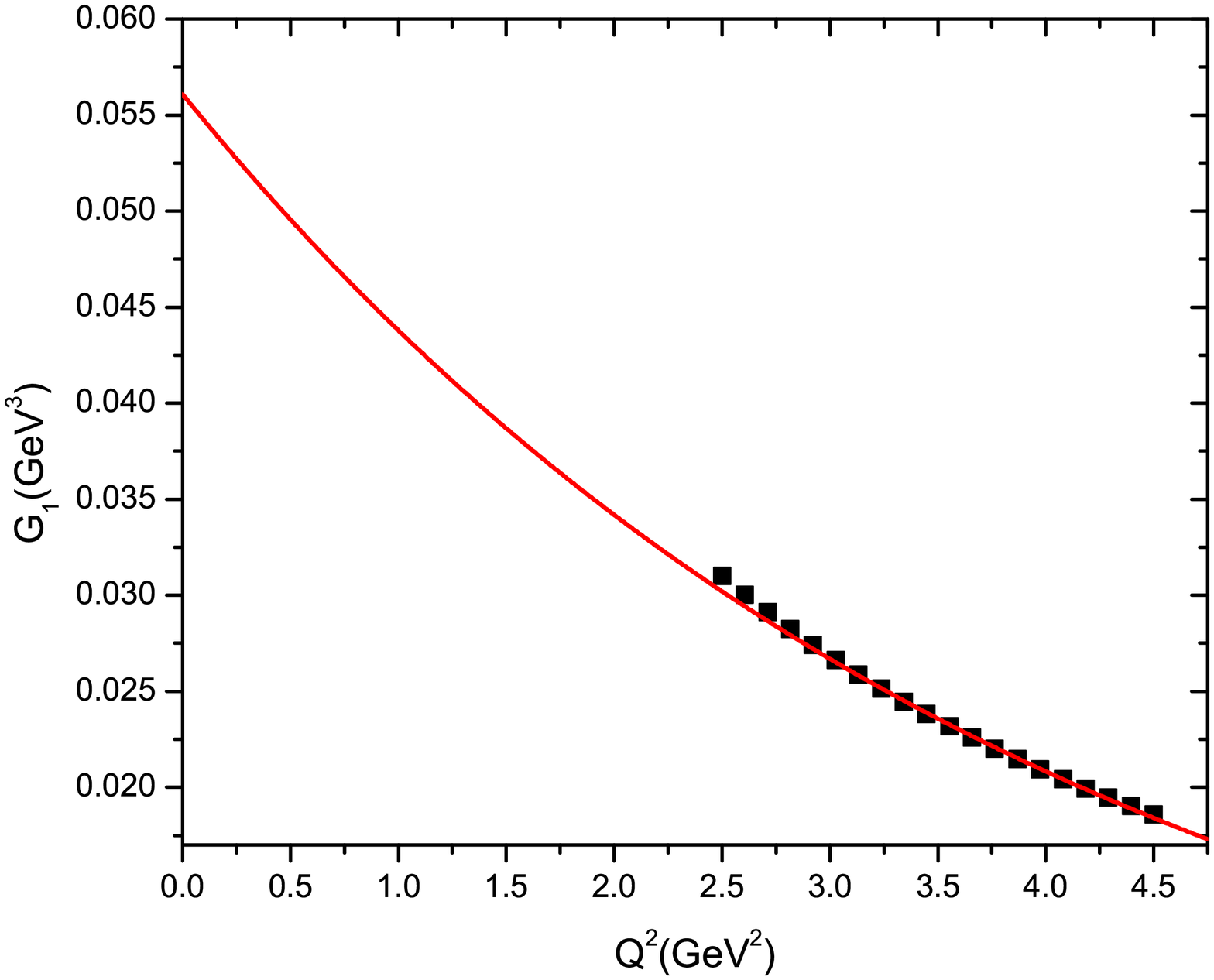}}\subfigure[]{\includegraphics[width=0.45\textwidth]{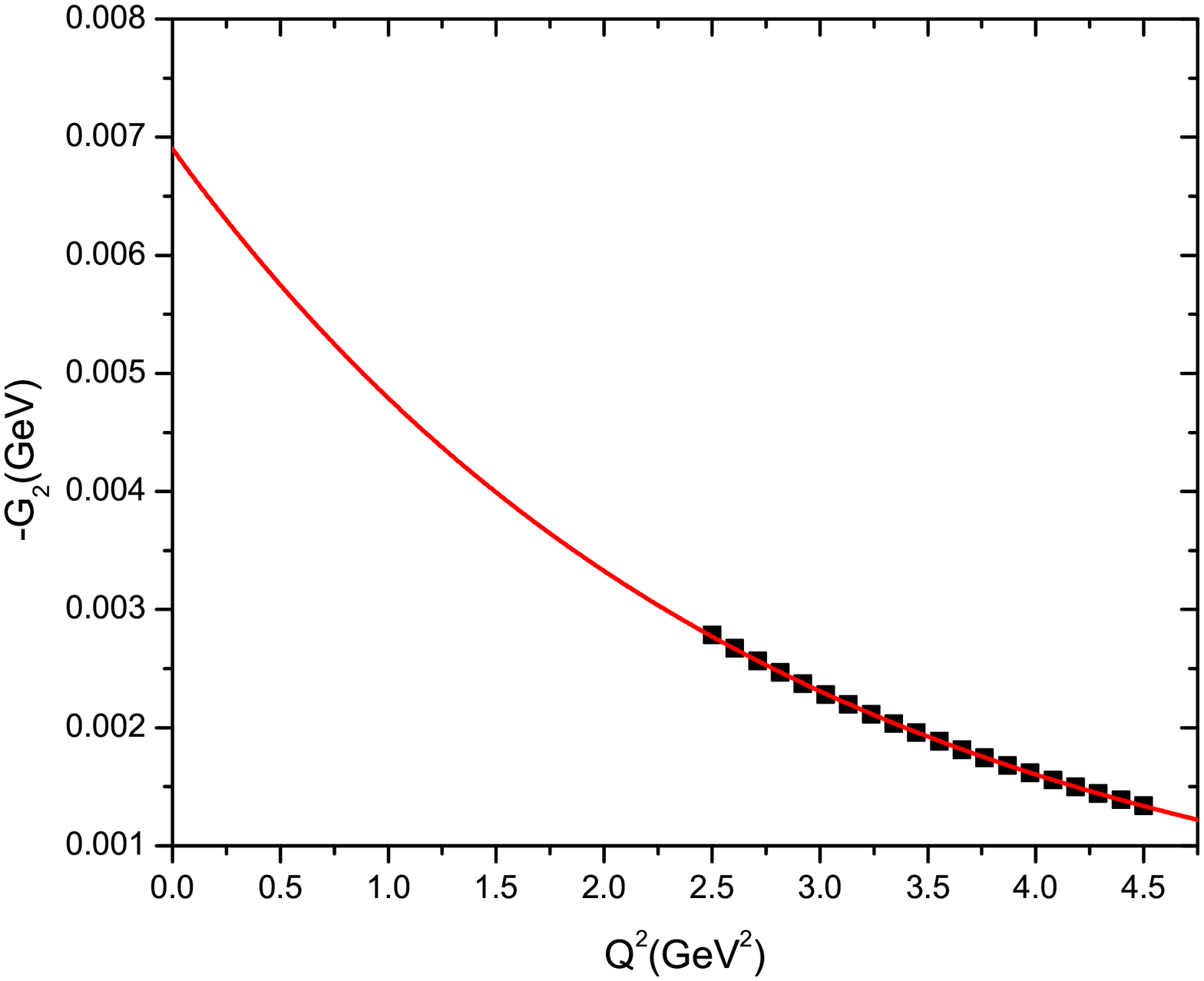}} 
\subfigure[]{\includegraphics[width=0.45\textwidth]{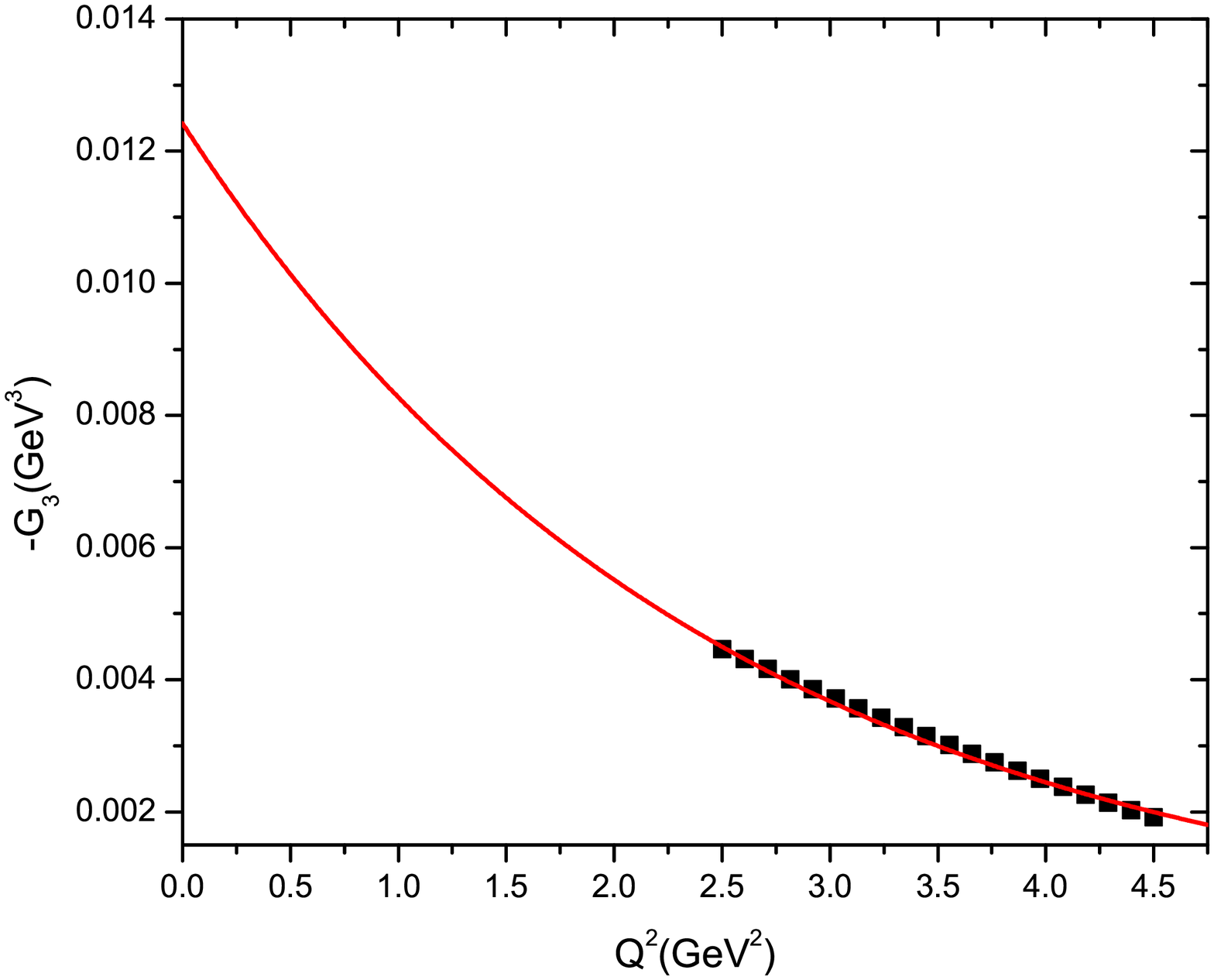}} 
  \caption{ \lb{fig3}   Momentum    dependence    of    the    functions     
for  $s_0^{1/2}=4.4\GeV$  and $u_0^{1/2}=3.6\GeV$: (a)  $G_1$, (b)   
$G_2$ and  (c) $G_3$. The solid line  gives the parametrization of the  
QCDSR results (dots) through Eq.~(\ref{giq2}) and the results in  
Table \ref{tab1}.} 
\end{figure*} 
In the  LHS of Eq.~(\ref{3sr}),  the unknown functions  $G_i(Q^2)$ and 
$H_i(Q^2)$ have  to be  determined by matching  both sides of  the sum 
rule.  In 
Fig.~\ref{fig2},  we  show  the  points  obtained if  we  isolate  the 
functions $G_i(Q^2)$ in Eq.~(\ref{3sr}) and vary both $Q^2$ and $M^2$. 
The  functions  $G_i(Q^2)$  [and consequently  $A(Q^2),B(Q^2),C(Q^2)$] 
should not depend on $M^2$, so we  limit our fit to a region where the 
function is clearly stable in $M^2$ to all values of $Q^2$. We can see 
in  Fig.~\ref{fig2}  that  the  regions  of  stability  in  $M^2$  for 
$G_1(Q^2)$  is  $7.0\GeV^2\leq  M^2\leq8.5\GeV^2$, for  $G_2(Q^2)$  is 
$6.5\GeV^2\leq M^2\leq7.5\GeV^2$, and for $G_3(Q^2)$ is $8.0\GeV^2\leq 
M^2\leq9.0\GeV^2$.

In Fig.~(\ref{fig3}) we show, through the dots, the QCDSR results 
for the functions   $G_i(Q^2)$ as a function of $Q^2$. The  form factors   
$A(Q^2),B(Q^2),$ and $C(Q^2)$  can be  easily obtained  by 
using      Eqs.(\ref{Aq2}), (\ref{Bq2}) and (\ref{Cq2}). Since the   
coupling constants, appearing in Eq.~(\ref{matrix}), are defined as   
the value of 
the form factors at the photon pole: $Q^2=0$, to determine the couplings 
$A$, $B$ and $C$ we have to extrapolate $A(Q^2),~B(Q^2)$, and $C(Q^2)$ to  
a region where the sum rules are no  longer valid (since  the QCDSR  
results  are valid at  the deep Euclidean region). To do that we fit 
the QCDSR results, shown in  Fig.~(\ref{fig3}), as  exponential functions: 
\beq  G_i(Q^2)=g_1 e^{-g_2  Q^2}\,.\lb{giq2}  
\enq  
 
We do  the fitting for $s_0^{1/2}=4.4\GeV$  and $u_0^{1/2}=3.6\GeV$ as 
the  results do  not depend  much  on this  parameters. The  numerical 
values of the fitting parameters are shown in the Table \ref{tab1}. 
 
\begin{table}[h] 
\begin{center} 
  \begin{tabular}{|c||c|c|c|}\hline\hline 
      &   $G_1$              &   $G_2$   & $G_3$ \\\hline\hline 
$g_1$   & $0.056\GeV^3$     &  $-0.0069\GeV$   &  $-0.013\GeV^3$\\ 
$g_2$   & $0.25\GeV^{-2}$    & $0.365\GeV^{-2}$ &  $0.41\GeV^{-2}$\\\hline 
\hline 
\end{tabular} 
\end{center} 
\caption{\lb{tab1} Results for the fitting parameters.} 
\end{table}

From Fig.~(\ref{fig3}) we can see  that the $Q^2$ dependence  of the 
QCDSR results for the functions   $G_i(Q^2)$    are   well   reproduced     
by   the   chosen parametrization,  in  the  interval  
$2.5\GeV^2\leq  Q^2\leq4.5\GeV^2$, where the QCDSR are valid.

Using Eqs.~(\ref{Aq2}),      (\ref{Bq2}),     (\ref{Cq2})      and 
(\ref{giq2}) and varying  $\theta$ in the range $5^o\leq\theta\leq13^o$ 
we get: 

\beqa 
A&=&A(Q^2=0)=18.65\pm0.94\,;\nn\\ 
A+B&=&(A+B)(Q^2=0)=-0.24\pm0.11\,;\nn\\ 
C&=&C(Q^2=0)=-0.843\pm0.008\,.\lb{resabc} 
\enqa

The decay width is given in terms of these couplings through \cite{dong08}: 
\beq 
\Gamma(X\to J/\psi~\gamma)=\frac{\alpha}{3}\frac{p^{*5}}{m_X^4} 
\bigg((A+B)^2+\frac{m_X^2}{m_\psi^2}(A+C)^2\bigg)\,,\lb{width}\nn\\ 
\enq   
where $p^*=(m_X^2-m_\psi^2)/(2m_X)$ is the three-momentum of the photon in the 
 $X$ rest frame. To compare  our results with the experimental  data shown in 
Eq.~(\ref{data}) we use the result  for the decay width of the channel 
$J/\psi\pi^+\pi^-$,  obtained  in   the  Ref.~\cite{x3872mix},  which  was 
computed in the same range  of the mixing angle $\theta$ and with 
the same angle $\alpha=20^0$: $\Gamma(X\to 
J/\psi~\pi\pi)=9.3\pm6.9\MeV$.  We get 
\beq 
\frac{\Gamma(X\to J/\psi~\gamma)}{\Gamma(X\to J/\psi~\pi^+\pi^-)}=0.19 
\pm0.13\,,\lb{brfinal} 
\enq 
which is in complete agreement with the experimental result.

%
%
\section{Conclusions} 
%
%
 
We have presented a QCDSR  analysis of the three-point function of the 
radiative  decay  of  the  $X(3872)$  meson by  considering  a  mixed 
charmonium-molecular current.   We find that the sum  rules results in 
Eqs.~(\ref{brfinal})  are compatible  with  experimental data.   These 
results   were  obtained   by   considering  the   mixing  angles   in 
Eq.~(\ref{4mix}) and  (\ref{field}) with the  values $\al=20^o$ and 
$5^\circ  \leq \theta  \leq  13^\circ$.  The  present  result is  also 
compatible with previous analysis of the mass of the $X$ state and 
the decays into  $J/\psi\pi^0\pi^+\pi^-$ and $J/\psi\pi^+\pi^-$ 
\cite{x3872mix}, since 
the  values of the  mixing angles  used in  both calculations  are the 
same. It is important to mention that there is no free parameter in the  
present analysis and, therefore, the result presented here strengthens  
the conclusion reached in Ref.~\cite{x3872mix} that the $X(3872)$ is  
probably a state with charmonium and molecular components.

\end{document}